\documentclass[twocolumn,pra,aps,superscriptaddress]{revtex4-1}
\usepackage{amssymb,amsmath,amsfonts}

\usepackage[latin9]{inputenc}
\usepackage{color}
\usepackage{amstext}
\usepackage{graphicx}
\usepackage[unicode=true,pdfusetitle,bookmarks=false,breaklinks=false,pdfborder={0 0 1},backref=false,colorlinks=false]{hyperref}
\hypersetup{bookmarksnumbered=false,bookmarksopen=false}

\usepackage{hyperref}
\hypersetup{colorlinks,citecolor=NavyBlue,filecolor=black,linkcolor=black,urlcolor=black}
\usepackage[dvipsnames]{xcolor}

\makeatletter

\newcommand\adag{a^\dagger}
\newcommand\adaga{a^\dagger a}

\newcommand\ket[1]{\left|#1\right\rangle}
\newcommand\bra[1]{\left\langle #1 \right|}

\usepackage{times}

\makeatother

\begin{document}
\title{Spin-boson model as a simulator of non-Markovian multiphoton Jaynes-Cummings models}

\author{R. Puebla}\email{r.puebla@qub.ac.uk}
\affiliation{Centre for Theoretical Atomic, Molecular, and Optical Physics, School of Mathematics and Physics, Queen's University, Belfast BT7 1NN, United Kingdom}

\author{G. Zicari}
\affiliation{Centre for Theoretical Atomic, Molecular, and Optical Physics, School of Mathematics and Physics, Queen's University, Belfast BT7 1NN, United Kingdom}

\author{I. Arrazola}
\affiliation{Department of Physical Chemistry, University of the Basque Country UPV/EHU, Apartado 644, E-48080 Bilbao, Spain}

\author{E. Solano}
\affiliation{Department of Physical Chemistry, University of the Basque Country UPV/EHU, Apartado 644, E-48080 Bilbao, Spain}
\affiliation{IKERBASQUE, Basque Foundation for Science, Maria Diaz de Haro 3, 48013 Bilbao, Spain}
\affiliation{Department of Physics, Shanghai University, 200444 Shanghai, China}

\author{M. Paternostro}
\affiliation{Centre for Theoretical Atomic, Molecular, and Optical Physics, School of Mathematics and Physics, Queen's University, Belfast BT7 1NN, United Kingdom}

\author{J. Casanova}
\affiliation{Department of Physical Chemistry, University of the Basque Country UPV/EHU, Apartado 644, E-48080 Bilbao, Spain}
\affiliation{IKERBASQUE, Basque Foundation for Science, Maria Diaz de Haro 3, 48013 Bilbao, Spain}

\begin{abstract}
  The paradigmatic spin--boson model considers a spin degree of freedom interacting with an environment typically constituted by a continuum of bosonic modes. This ubiquitous model is of relevance in a number of physical systems where, in general, one has neither control over the bosonic modes, nor the ability to tune distinct interaction mechanisms. Despite this apparent lack of control, we present a suitable transformation that approximately maps the spin-boson dynamics into that of a {\it tunable} multiphoton Jaynes-Cummings model undergoing dissipation. Interestingly, the latter model describes the coherent interaction between a spin and a single bosonic mode via the simultaneous exchange of {\it n} bosons per spin excitation. Resorting to the so-called reaction coordinate method, we identify a relevant collective bosonic mode in the environment, which is then used to generate multiphoton interactions following the proposed theoretical framework. Moreover, we show that spin-boson models featuring structured environments can lead to non-Markovian multiphoton Jaynes-Cummings dynamics. We discuss the validity of the proposed method depending on the parameters and analyse its performance, which is supported by numerical simulations. In this manner, the spin-boson model serves as a good analogue quantum simulator for the inspection and realization of multiphoton Jaynes-Cummings models, as well as the interplay of non-Markovian effects and, thus, as a simulator of light-matter systems with tunable interaction mechanisms.
  \end{abstract}

\maketitle

\section{Introduction}

The rapid technological progress we have experienced during the last few decades has made possible previously inconceivable experiments at the quantum regime, boosting their degree of precision, isolation and control to unprecedented limits~\cite{Dowling:03}. Currently, quantum systems can be inspected in a very controllable manner in a number of distinct setups. This experimental breakthrough has therefore stimulated the emergence of research areas such as quantum information and computation and quantum simulation, where the exploitation of quantum effects will allow us to surpass both the capabilities of their classical counterparts in the near future~\cite{Nielsen}.
In particular, quantum simulation considers a scenario in which a well-controlled quantum system serves as a simulator of other inaccessible systems~\cite{Feynman:82,Johnson:14,Georgescu:14}. In this manner, interesting quantum dynamics (i.e., the target dynamics) may be explored using, for example, optical lattices~\cite{Bloch:12} or trapped ions~\cite{Blatt:12}. The target dynamics can be obtained either by decomposing the time-evolution propagator in a set of simple quantum operations (digital quantum simulation) or by finding a map that brings the Hamiltonian into the desired form of the model to be simulated (analogous to quantum simulation)~\cite{Georgescu:14}.
In this article, we will consider the latter method, by using as a quantum simulator the paradigmatic spin-boson model~\cite{Leggett:87,Weiss}.

The spin-boson model describes a spin immersed in an environment formed by a large, typically infinite, number of bosonic modes, in contrast to the quantum Rabi or Jaynes-Cummings models where the interaction comprises a single bosonic mode~\cite{Rabi:36,Rabi:37,Jaynes:63,Scully}. The spin-boson model encompasses very rich physics depending on how the spin couples with the distinct bosonic modes. 
Hence, while it is a minimal model to scrutinize the quantum effects of dissipation, it has application in a broad range of systems~\cite{Weiss,Leggett:87}, ranging from defects in solid state platforms to quantum emitters in biological systems~\cite{Huelga:13}. Moreover, much effort inspecting the spin-boson model has dealt with its critical behaviour, that is with the emergence of a quantum phase transition between a delocalized and a localized phase of the spin degree of freedom as one increases the spin-environment coupling~\cite{Leggett:87}. The simulation of the spin-boson model {(or of a generic open quantum system)} in the strong coupling regime is however computationally very demanding, as acknowledged in~\cite{Tanimura:89,Tanimura:90,Prior:10,Dattani:12,Dattani:13,Wilkins:15,Strathearn:18}, since the spin and the bosonic modes become entangled, forming a truly quantum many-body system. In some situations, one can still resort to analytical methods, which may simplify the problem considerably. Among these methods one finds the so-called reaction coordinate mapping~\cite{Thoss:01,Martinazzo:11,Iles:14,Iles:16,Strasberg:16,Strasberg:18,Nazir:18}, which can be viewed as a first step of the more general semi-infinite chain mapping of the environmental degrees of freedom~\cite{Chin:10,Woods:14}. The reaction coordinate is defined as a collective mode of the original environment oscillators. In this manner, one can bring the spin-boson model into the form of a generalized quantum Rabi model~\cite{Rabi:36,Rabi:37,Scully} whose bosonic mode undergoes dissipation as it interacts with the residual environment. In particular cases, upon rearranging the original environmental degrees of freedom, the dissipation acquires a Markovian character, hence simplifying considerably the complexity of the problem (see for example~\cite{Iles:14}). It is also worth mentioning other attempts to capture quantum dynamics effectively with complex system-environment interactions, as for example the recent work relying on pseudo-modes~\cite{Mascherpa:19}, which builds on the proven equivalence for the dynamics of the system in both frames~\cite{Tamascelli:18}.

The quantum Rabi model (QRM), as well as its simplified version known as the Jaynes-Cummings model (JCM)~\cite{Jaynes:63} play a central role in the description of light-matter interacting systems and in quantum information science~\cite{Scully,Nielsen}. In these models, the interaction mechanism between the spin and bosonic degrees of freedom has a linear form, namely the spin gets excited or deexcited by absorbing or emitting one bosonic excitation. While this interaction is ubiquitous in quantum physics and with application in various experimental platforms~\cite{Braak:16}, other forms of a spin-boson exchange mechanisms beyond this simple case are also of interest.
On the one hand, interactions beyond the linear fashion are of relevance for several applications in quantum computation and simulation (e.g., the Kerr effect~\cite{Lloyd:99}). Furthermore, these exchange mechanisms may unveil interesting phenomena in light-matter systems~\cite{Felicetti:15,Pedernales:18}, as well as in their multiple spin counterparts~\cite{Garbe:17}. One possible generalization of the QRM or JCM consists of considering a spin-multiphoton interaction, where the spin exchanges $n$ excitations simultaneously with the bosonic mode. Such a generalization is often regarded as $n$-photon QRM or JCM, (nQRM or nJCM), and it has recently attracted attention mainly in its $n=2$ form~\cite{Felicetti:15,Puebla:17pra,Cui:17,Pedernales:18,Felicetti:18,Xie:19}, although models with $n>2$ have been also analysed~\cite{Lo:98}. From an experimental point of view, however, such multiphoton terms are typically hard to attain. Thus, its realization may benefit from quantum simulation protocols, allowing for enough tunability and control over multiphoton interaction terms, as proposed using optical trapped ions~\cite{Felicetti:15,Puebla:17pra} or superconducting qubits~\cite{Felicetti:18}. These latter schemes realize effective multiphoton exchange terms by exploiting the nonlinear fashion in which the spin and bosonic degrees of freedom couple. It is however still possible to realize such multiphoton models even when the setup comprises solely a linear, i.e., standard, interaction mechanism, and thus, it is not suited for a direct simulation of these models, as shown in~\cite{Casanova:18}.

In this article, we follow the theoretical framework developed in~\cite{Casanova:18,Puebla:19}, combining the ideas of the reaction-coordinate mapping~\cite{Thoss:01,Martinazzo:11,Iles:14,Iles:16,Strasberg:16,Strasberg:18,Nazir:18} {to show that the paradigmatic spin-boson model, featuring a continuum of bosonic modes, can serve as an analogue quantum simulator for the realization of different dissipative multiphoton Jaynes-Cummings models by tuning the frequency and bias parameter of the spin. In this manner, we demonstrate the emergence of a connection between the dynamics of these paradigmatic and fundamental quantum models, which was not previously unveiled. Moreover, as the spin-boson model is of considerable experimental significance, i.e., it describes the ubiquitous scenario of a two-level system interacting with an arbitrary environment, our method paves the way for the simulation of multiphoton Jaynes-Cummings models in distinct setups.}
In particular, by considering a full spin-boson model, we naturally extend the theoretical framework beyond the standard local master equation description of dissipation effects in the simulator, as considered in~\cite{Puebla:19}. 
Furthermore, we show that the simulated multiphoton Jaynes-Cummings models may acquire non-Markovian behaviours when the spin-boson model features a structured environment, thus highlighting the suitability of the proposed theoretical framework to explore aspects of non-Markovianity in distinct light-matter interacting systems.

The article is organized as follows. In Section~\ref{s:sbm}, we introduce the spin-boson model, while in Section~\ref{s:theory}, we explain how to map the spin-boson model into a different Hamiltonian comprising the desired spin-multiphoton interaction terms and discuss how the dissipative effects must be transformed {into} the aimed model. For that, we first introduce the reaction coordinate mapping in Section~\ref{ss:rc}, while in Section~\ref{ss:senv}, we explain how to extend the theoretical framework to incorporate further bosonic modes in the realization of the desired multiphoton model. After having provided the theoretical derivation of how to perform the analogue quantum simulation, we present examples and numerical results for the simulation of different multiphoton Jaynes-Cummings models in Section~\ref{s:num}. Finally, we summarize the main conclusions of this article in Section~\ref{s:conc}.

\section{The spin-boson model}\label{s:sbm}
The spin-boson model describes a two-level system interacting with a large, typically infinite, number of bosonic modes, which constitute the environment. This model has been acknowledged as a paradigm for the inspection of quantum dissipation and quantum-to-classical transition~\cite{Weiss,Leggett:87}. As many physical systems can be well approximated as a two-level system for sufficient low temperature, the spin-boson model has become a cornerstone in the description of quantum effects in diverse physical realizations, ranging from quantum-based setups~\cite{Leggett:87,Weiss} to biological complexes~\cite{Huelga:13}. In addition, this model has played a key role in the development of the theory of open quantum systems~\cite{Breuer}, providing a suitable test-bed to benchmark distinct approximations and tools aimed to deal with the large number of environment degrees of freedom efficiently. Moreover, the relevance of the spin-boson model also encompasses the context of critical systems, as it features a quantum phase transition between spin localized and delocalized phases (see Refs.~\cite{Vojta:06,Hur:10} and the references therein). Hence, the spin--boson model exhibits rich physics, and it is of fundamental relevance in many different areas of research.

\begin{figure} 
\centering
\includegraphics[width=1.\linewidth,angle=00]{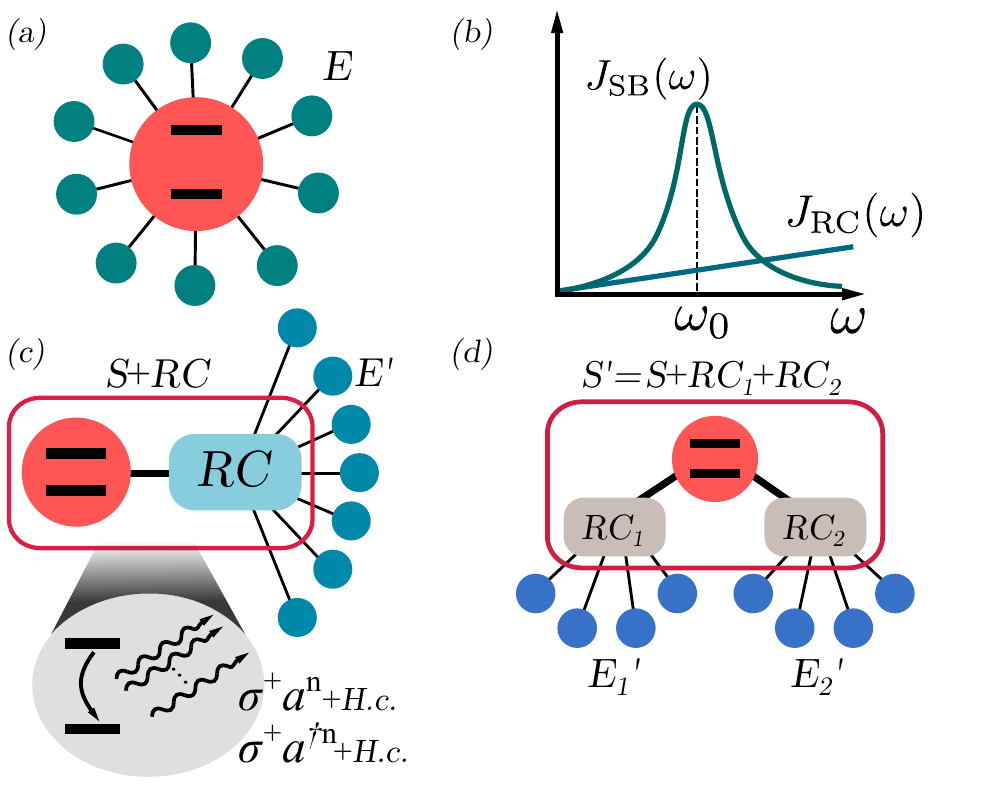}
\caption{\small{{{(a)} Spin-boson model in the customary star configuration, }
where each of the circles corresponds to a harmonic oscillator of the environment with frequency $\omega_k$ interacting with the spin through $\sigma_x f_k(c_k+c_k^{\dagger})$, before the reaction coordinate mapping. In {(b)}, we show an underdamped spin-boson spectral density $J_{\rm SB}(\omega)$, peaked at $\omega_0$ (cf. Equation~(\ref{eq:JSBUD})). Upon the reaction coordinate mapping, a collective degree of freedom is included into the system, which in turn interacts with the residual environment, as sketched in {(c)} (see the main text for further details). For an underdamped $J_{\rm SB}(\omega)$, $J_{\rm RC}(\omega)$ adopts a Markovian form, as depicted in {(b)}. Such interaction with a collective coordinate can be exploited to realize Hamiltonians containing multiphoton interaction terms, as indicated in {(c)} and explained in detail in Section~\ref{s:theory}. For structured environments, one can still rearrange the original environment using more collective coordinates into the augmented system $S'$, where each of them interacts now with its own residual environment, as sketched in {(d)} (see~\ref{ss:senv} for further details).}}
\label{fig:scheme}

\end{figure}

The Hamiltonian of the spin-boson model can be written as:
\begin{equation}\label{eq:sbm}
H_{\rm SB}=H_{\rm S}+H_{\rm E}+H_{\rm S-E}
\end{equation}
where each contribution reads as:
\begin{align}\label{eq:sbmHs}
 H_{\rm S}&=\frac{\epsilon_0}{2}\sigma_z+\frac{\Delta_0}{2}\sigma_x,\\
 H_{\rm E}&=\sum_k \omega_k c_k^{\dagger}c_k,\\ \label{eq:sbmHSE}
 H_{\rm S-E}&=\sigma_x\sum_k f_k (c_k+c_k^{\dagger}).
\end{align}
The first two terms represent the free-energy Hamiltonians of the spin and environment, while the last describes the interaction between them. Here, we consider that the frequency splitting of the spin is given by $\Delta_0$, while $\epsilon_0$ accounts for the bias between the eigenstates of the two-level system $\ket{\pm}$ and with $\vec{\sigma}=(\sigma_x,\sigma_y,\sigma_z)$ the usual spin-$\frac{1}{2}$ Pauli matrices (see Figure~\ref{fig:scheme}{(a)}). Hence, $\sigma_x\ket{\pm}=\pm \ket{\pm}$, $\sigma_z\ket{e}=\ket{e}$ and $\sigma_z\ket{g}=-\ket{g}$. The interaction with the environment is dictated by $H_{\rm S-E}$, where the $k^{\text{th}}$ mode with energy $\omega_k$ is coupled to the spin with a strength $f_k$. These bosonic modes fulfil the usual commutation relation $[c_k,c_{k'}^{\dagger}]=\delta_{k,k'}$. {Remarkably, the system--environment interaction can be completely characterized in terms of the spectral density, $J_{\rm SB}(\omega)=\sum_k f_k^2\delta(\omega-\omega_k)$, which here is assumed to be known. In anticipation of the developed theoretical framework that allows us to bring $H_{\rm SB}$ into the form of a multiphoton Jaynes--Cummings model, we comment that while the frequency splitting $\Delta_0$ tunes the multiphoton order of the interaction, the bias parameter $\epsilon_0$ will be proportional to the interacting strength of the simulated model (see Section~\ref{s:theory}).}

{In addition, we comment that one could consider the application of $n_d$ drivings onto the spin. As discussed in~\cite{Casanova:18,Puebla:19}, under certain conditions that we will explain in the following section, applying spin drivings enables the simultaneous realization of different multiphoton Jaynes-Cummings interaction terms. In this manner, while a multiphoton Jaynes-Cummings model can be attained without the need for any driving, $n_d=0$, the realization of a multiphoton quantum Rabi model requires the application of at least one, i.e., $n_d=1$.} In general, the free-energy Hamiltonian of the spin under $n_d$ drivings with amplitude $\epsilon_j$ and detuning $\Delta_j$ with respect to the spin frequency splitting $\Delta_0$ reads as:
\begin{align}\label{eq:sbmtime}
  H_{\rm S,d}=\frac{\Delta_0}{2}\sigma_x&+\sum_{j=0}^{n_d} \frac{\epsilon_j}{2} \left[\cos(\Delta_j-\Delta_0)t\ \sigma_z\right.\nonumber\\&+\left.\sin(\Delta_j-\Delta_0)t\ \sigma_y\right].
\end{align}
Clearly, setting $\epsilon_{j>0}=0$ (or $\Delta_{j}=\Delta_0$), we recover the form of the standard drivingless $H_{\rm S}$ given in Equation~(\ref{eq:sbmHs}). For the sake of simplicity, in this article, we will focus on cases with $n_d=0$, i.e., aiming to realize multiphoton Jaynes-Cummings models. However, we stress that the procedure explained in the following can be applied in a straightforward manner when $n_d>0$.

\section{Analogue simulation of multiphoton spin--boson interactions}\label{s:theory}

The task now consists of bringing the spin-boson Hamiltonian $H_{\rm SB}$ into the form of a $n$-photon model, i.e, into a model containing interaction terms of the form $\sigma^{\pm}a^{n}$ and $\sigma^{\pm}(\adag)^n$. For that, one could perform the approximate mapping used in~\cite{Casanova:18,Puebla:19} directly onto $H_{\rm SB}$. This would require the selection of a particular bosonic mode out of the environment with frequency $\omega_q$ to now play the role of $a$ in the interaction with the spin ($c_q\rightarrow a$), while treating the rest of $c_{k\neq q}$ as a residual environment. Here, however, we resort to a more sophisticated procedure, based on the so-called reaction coordinate (RC) mapping~\cite{Thoss:01,Martinazzo:11,Iles:14,Iles:16,Strasberg:16,Strasberg:18,Nazir:18}, which consists of rearranging the environment degrees of freedom, such that a small number of collective coordinates can be included in the Hamiltonian part, which in turn interact with the residual environment. In certain cases, the open-quantum system description of the augmented system is considerably simplified with respect to the original system plus environment. Clearly, if the spin--boson model involves just a discrete number of modes, the reaction-coordinate procedure then trivially retrieves the original discrete environment. 

\subsection{Reaction coordinate mapping}\label{ss:rc}
In the following, we summarize how to make use of the RC mapping for a spin-boson model, which has been studied previously in different works~\cite{Iles:14,Iles:16}, while referring to Appendix~\ref{app:a} and References~\cite{Thoss:01,Martinazzo:11,Iles:14,Iles:16,Strasberg:16,Strasberg:18,Nazir:18} for further details of the calculations and of the RC mapping.

We shall start by defining a collective mode or reaction coordinate, described by the annihilation and creation operators $a$ and $\adag$, such that:
\begin{align}
\lambda(a+\adag)=\sum_k f_k(c_k+c^{\dagger}_k),
\end{align}
while the residual environmental degrees of freedom transform into $b_k$ and $b^{\dagger}_k$, requiring that the latter appear in a normal form in the Hamiltonian. In this manner, the original spin-boson Hamiltonian adopts the form of $H_{\rm SB}=H_{\rm S+RC}+H_{\rm RC-E'}+H_{\rm E'}$, where the former is given by:
\begin{align}\label{eq:HSRC}
&H_{\rm S+RC}=\frac{\Delta_0}{2}\sigma_x+\Omega\adaga+\lambda\sigma_x(a+\adag)\nonumber\\&\quad+\sum_{j=0}^{n_d}\frac{\epsilon_j}{2}\left[\cos(\Delta_j-\Delta_0)t\ \sigma_z+\sin(\Delta_j-\Delta_0)t\ \sigma_y \right],
\end{align}
and the other two terms are $H_{\rm RC-E'}+H_{\rm E'}=(a+\adag)\sum_k g_k(b_k+b_k^{\dagger})+(a+\adag)^2\sum_k\frac{g_k^2}{\omega_k}+\sum_k \omega_kb_k^{\dagger}b_k$. The reaction coordinate map is completed upon the identification of the parameters $\lambda$, $\Omega$, and $g_k$ or, thus, $J_{\rm RC}(\omega)=\sum_k g_k^2\delta(\omega-\omega_k)$. For certain cases, such mapping allow for an exact relation between the original and transformed parameters~\cite{Nazir:18}. Indeed, considering an underdamped spin-boson spectral density in the initial spin-boson model,
\begin{align}\label{eq:JSBUD}
 J_{\rm SB}(\omega)=\frac{\alpha\Gamma\omega_0^2\omega}{(\omega_0^2-\omega^2)^2+\Gamma^2\omega^2},
\end{align}
one can show that the resulting spectral density for the residual environment interacting with the reaction coordinate reads as:
\begin{align}
 J_{\rm RC}(\omega)=\gamma\omega e^{-\omega/\Lambda}
\end{align}
provided $\Lambda/\omega\gg 1$ and where the parameters are related according to $\gamma=\Gamma/(2\pi\omega_0)$, $\Omega=\omega_0$, and $\lambda=\sqrt{\pi\alpha\omega_0/2}$ (see Appendix~\ref{app:a} or~\cite{Thoss:01,Martinazzo:11,Iles:14,Nazir:18} for further details of this derivation). Here, the frequency $\omega_0$ in $J_{\rm SB}(\omega)$ denotes the position at which the spectral density features a maximum, while $\Gamma$ and $\alpha$ account for its width and strength, respectively. For $J_{\rm RC}(\omega)$, the coupling strength is given by $\gamma$. In this manner, by augmenting the system incorporating a collective mode, the original spin--boson model with $J_{\rm SB}(\omega)$ is transformed into a spin plus reaction coordinate, which now in turn interacts with a Markovian environment, where the standard Born--Markov approximations can be performed~\cite{Breuer}. Indeed, the master equation governing the dynamics of the augmented system, spin plus reaction coordinate, reads as (see Appendix~\ref{app:a} for the details of the calculation, which closely follows~\cite{Iles:14}):
\begin{align}\label{eq:mesrc}
\dot{\rho}_{\rm S+RC}(t)=&-i\left[H_{\rm S+RC},\rho_{\rm S+RC}(t)\right]-\left[x,\left[\chi,\rho_{\rm S+RC}(t) \right] \right]\nonumber\\&+\left[x,\left\{\Theta,\rho_{\rm S+RC}(t) \right\} \right].
\end{align}
with $x=a+\adag$, while the quantities $\chi$ and $\Theta$ define the rates affecting the reaction coordinate. They are {defined} as:
\begin{align}\label{eq:chi}
 \chi&\approx \frac{\pi}{2}\sum_{jk} J_{\rm RC}(\xi_{jk})\coth(\beta\xi_{jk}/2)x_{jk}\ket{\phi_j}\bra{\phi_k},\\ \label{eq:theta}
 \Theta&\approx\frac{\pi}{2}\sum_{jk} J_{\rm RC}(\xi_{jk})x_{jk}\ket{\phi_j}\bra{\phi_k},
\end{align}
where $x_{jk}=\bra{\phi_j}x\ket{\phi_k}$, $H_{\rm S+RC}\ket{\phi_j}=\varphi_j\ket{\phi_j}$, and $\xi_{jk}=\varphi_j-\varphi_k$.

Having obtained the reaction coordinate Hamiltonian, we undertake the transformation of $H_{\rm S+RC}$, and thus, of Equation~(\ref{eq:mesrc}), to achieve a model that comprises spin-multiphoton interaction terms. For that purpose, we will introduce two auxiliary Hamiltonians $H_{a}$ and $H_{b}$, which will arise in the intermediate steps by moving into a suitable interaction picture and transforming them accordingly. The first step consists indeed of moving to a rotating frame in which $H_{\rm S+RC}\equiv H_{a,1}^I$ where $H_{a}=H_{a,0}+H_{a,1}$ with $H_{a,0}=-\Delta_0/2\sigma_x$. In this manner, we find:
\begin{align}
 H_{a}=\Omega\adaga&+\lambda\sigma_x(a+\adag)\nonumber\\&+\sum_{j=0}^{n_d}\frac{\epsilon_j}{2}\left[\cos\Delta_j t \sigma_z+\sin\Delta_j t\sigma_y \right].
\end{align}
while Equation~(\ref{eq:mesrc}) transforms into:
\begin{equation}\label{eq:merhoa}
\dot{\rho}_a(t){=}-i\left[H_{a},\rho_{a}(t)\right]-[x,[\hat{\chi},\rho_{a}(t) ] ]+[x,\{\hat{\Theta},\rho_{a}(t) \} ].
\end{equation}
where $\hat{\chi}=U_{a,0}\chi U_{a,0}^{\dagger}$ and $\hat{\Theta}=U_{a,0}\Theta U_{a,0}^{\dagger}$, such that $U_x=\mathcal{T}e^{-i\int_0^tds H_x(s)}$ is the time-evolution operator of a Hamiltonian $H_x$. Then, we perform a further transformation using the unitary operator $T(\alpha)$, defined as $T(\alpha)=1/\sqrt{2}\left[D^{\dagger}(\alpha)\left(\ket{e}\bra{e}-\ket{g}\bra{e}\right)+D(\alpha)\left(\ket{g}\bra{g}+\ket{e}\bra{g}\right)\right]$ with $D(\alpha)=e^{\alpha \adag-\alpha^*a}$ the standard displacement operator. Hence, $H_b\equiv T^{\dagger}(-\lambda/\Omega)H_a T(-\lambda/\Omega)$ such that $\rho_b=T^{\dagger}\rho_aT$, which leads to (see Appendix~\ref{app:b} for further details):
\begin{align}
\dot{\rho}_b=-i\left[H_b,\rho_b\right]&-\left[T^{\dagger}xT,\left[T^{\dagger}\hat{\chi}T,\rho_{b}(t) \right] \right]\nonumber\\&+\left[T^{\dagger}xT,\left\{T^{\dagger}\hat{\Theta}T,\rho_{b}(t) \right\} \right],
\end{align}
where the Hamiltonian $H_b$ can be written as:
\begin{align}\label{eq:Hb}
H_b=\Omega\adaga+\sum_{j=0}^{n_d}\frac{\epsilon_j}{2}\left[ \sigma^+e^{2\lambda(a-\adag)/\Omega}e^{-i\Delta_jt}+{\rm H.c.}\right].
\end{align}
Hence, the dissipator acting on $\rho_b$ has the same form as in Equation~(\ref{eq:merhoa}), but with transformed operators, namely $T^{\dagger}xT$, $T^{\dagger}\hat{\chi}T$, and $T^{\dagger}\hat{\Theta}T$, where $T\equiv T(-\lambda/\Omega)$. Finally, by moving to an interaction picture with respect to $H_{b,0}=(\Omega-\tilde\nu)\adaga-\tilde\omega\sigma_z/2$ and expanding the exponential in Equation~(\ref{eq:Hb}) (the latter requires that $|2\lambda/\Omega|\sqrt{\left<(a+\adag)^2\right>}\ll 1$ for truncating the exponential to a finite number of terms), we arrive at a Hamiltonian containing multiphoton interaction terms. The latter condition is commonly known as the Lamb-Dicke regime. In addition, we consider the driving frequencies to be $\Delta_{j}=\pm n_j (\tilde\nu-\Omega)-\tilde\omega$ with $|\Omega-\tilde{\nu}|\gg \epsilon_j/2$, so that one can safely perform a rotating-wave approximation keeping only those terms that are resonant, i.e., time independent (see Appendix~\ref{app:b} for further details of the calculation). Note that, as $H_b$ is similar to the Hamiltonian describing an optical trapped ion under the action of lasers driving vibrational sidebands~\cite{Leibfried:03}, the procedure to obtain Jaynes-Cummings or quantum Rabi models is analogous to those cases~\cite{Pedernales:15,Felicetti:15,Lv:18}. In this manner, we can approximate $H_{b,1}^I\equiv U_{b,0}^{\dagger}H_{b,1}U_{b,0}\approx H_{\rm n}$, where $H_{\rm n}$ contains the aimed at multiphoton interactions,
\begin{align}\label{eq:Hn}
H_{\rm n}=\frac{\tilde\omega}{2}\sigma_z&+\tilde\nu\adaga+\sum_{j\in r}\frac{\epsilon_j(2\lambda)^{n_j}}{2\Omega^{n_j}n_j!}\left[\sigma^+a^{n_j}+{\rm H.c.}\right]\nonumber\\&+\sum_{j\in b}\frac{\epsilon_j(2\lambda)^{n_j}}{2\Omega^{n_j}n_j!}\left[\sigma^+(-\adag)^{n_j}+{\rm H.c.}\right].
\end{align}
Note that the sets $r$ and $b$ encompass the terms with amplitude $\epsilon_j$ driving red- and blue-sidebands, that is those terms in Equation~(\ref{eq:sbmtime}) with frequency $\Delta_{j\in r}=+n_j(\tilde{\nu}-\Omega)-\tilde{\omega}$ and $\Delta_{j\in b}=-n_j(\tilde{\nu}-\Omega)-\tilde{\omega}$. Each of these drivings will contribute with a multiphoton interaction, either $\sigma^+a^{n_j}+{\rm H.c.}$ for $j\in r$ or $\sigma^-a^{n_j}+{\rm H.c.}$ for $j\in b$, which produce transitions between the states $\ket{m}\ket{g}\leftrightarrow \ket{m\mp n_j}\ket{e}$.{We stress that for a time-independent spin-boson model, as given in Equations~(\ref{eq:sbm})--(\ref{eq:sbmHSE}) (or equivalently with $n_d=0$ in $H_{S,d}$ as given in Equation~(\ref{eq:sbmtime}), one obtains a single $n$-photon [anti]
-Jaynes--Cummings interaction term, $\sigma^+a^{n}+{\rm H.c.}$ [$\sigma^+(-\adag)^{n}+{\rm H.c.}$], by choosing $\Delta_0=n(\tilde{\nu}-\Omega)-\tilde{\omega}$ [$\Delta_0=-n(\tilde{\nu}-\Omega)-\tilde{\omega}$] in the original spin--boson Hamiltonian $H_{\rm SB}$. Thus, one needs the knowledge of the relevant bosonic frequency $\Omega$ to simulate multiphoton interaction terms properly.}

In order to show how the dissipative part transforms, it is advisable to introduce the time-dependent unitary operator:
\begin{align}
 \Phi=U_{b,0}^{\dagger}T^{\dagger}U_{a,0}.
\end{align}
Then, one can see that, defining $\tilde\chi=\Phi \chi \Phi^{\dagger}$, $\tilde\Theta=\Phi \Theta \Phi^{\dagger}$ and $\tilde{x}=\Phi(a+\adag)\Phi^{\dagger}$, the resulting master equation for $\rho_{\rm n}(t)$ is:
\begin{equation}\label{eq:merhon}
\dot{\rho}_{\rm n}(t)=-i[H_{\rm n},\rho_{\rm n}(t)]-\left[\tilde{x},\left[\tilde\chi,\rho_{\rm n}(t)\right]\right]+[\tilde{x},\{\tilde\Theta,\rho_{\rm n}(t)\}]
\end{equation}
where the state $\rho_{\rm n}(t)$ of the multiphoton model is related to the original spin-boson upon the reaction coordinate mapping, $\rho_{\rm S+RC}(t)$, through a unitary transformation:
\begin{align}\label{eq:rhonSRC}
 \rho_{\rm n}(t)\approx \Phi\rho_{\rm S+RC}(t)\Phi^{\dagger}.
\end{align}
From the previous expression, it follows that the purity of the total state $\rho_{\rm S+RC}$ and that of $\rho_{\rm n}$ are approximately equal. Moreover, the reduced spin state in the different frameworks are related according to ${\rm Tr}_{\rm B}[\rho_{\rm SB}(t)]={\rm Tr}_{\rm RC}[\rho_{\rm S+RC}(t)]\approx {\rm Tr}_{\rm RC}[\Phi^{\dagger} \rho_{\rm n}(t)\Phi]$, where ${\rm Tr}_{\rm B}[\cdot]$ and ${\rm Tr}_{\rm RC}[\cdot]$ denote the trace over the environment degrees of freedom and reaction coordinate, respectively. In this manner, having access to the spin degree of freedom, one can have access to the dissipative spin dynamics dictated by the master equation~(\ref{eq:merhon}) under a multiphoton Hamiltonian $H_{\rm n}$, given in Equation~(\ref{eq:Hn}), whose parameters can be tuned. In addition, we remark that the initial state at $t_0=0$ in the multiphoton frame is related to that of the spin--boson model as $\rho_{\rm n}(0)=T^{\dagger}\rho_{\rm S+RC}(0)T$.

At this stage, a few comments regarding the validity of Equation~(\ref{eq:rhonSRC}) are in order. While the steps performed from $H_{\rm S+RC}$ to $H_{b}$ are exact, $H_{\rm n}$ is attained in an approximate manner. The good functioning of the simulation depends on how these approximations are met. That is, Equation~(\ref{eq:rhonSRC}) holds within the Lamb-Dicke regime $|2\lambda/\Omega|\sqrt{\left<(a+\adag)^2\right>}\ll 1$ and for parameters satisfying $|\Omega-\tilde\nu|\gg \epsilon_j/2 \ \forall j$, so that one can perform a rotating-wave approximation. {As a consequence, this approximation also sets a constraint on the total duration for a good simulation (see Appendix~\ref{app:b}).} Note that, as the parameters $\lambda$ and $\Omega$ are directly related to the original spin-boson spectral density, these conditions set constraints onto the accessible parameters, as well as on the temperature of the environment. Furthermore, in order to observe coherent multiphoton dynamics, the noise rates in Equation~(\ref{eq:merhon}) must be small compared to the parameters involved in $H_{\rm n}$. For the considered shape of $J_{\rm SB}(\omega)$, this translates into $\Gamma\ll \tilde\nu,\tilde{g}_n$, where $\tilde{g}_n=\epsilon_0 (2\lambda)^{n}/(2\Omega^nn!)$ for an $n_d=0$ and $\Delta_0=\pm n(\tilde\nu-\Omega)-\tilde\omega$ (cf. Equation~(\ref{eq:Hn}).

Finally, we comment that the previous scheme can be carried out beyond the Lamb-Dicke regime~\cite{Puebla:19}. Admittedly, when the Lamb-Dicke approximation does not hold, the Hamiltonian $H_{\rm n}$ is no longer a good approximation to the dynamics. In this case, the Hamiltonian $H_{\rm n}$ must be replaced by a suitable nonlinear Jaynes-Cummings or quantum Rabi model, whose coupling constants crucially depend on the Fock-state occupation number in a nonlinear fashion~\cite{MatosFilho:94,MatosFilho:96,Vogel:95,Cheng:18}. These nonlinear, yet multiphoton Hamiltonians appear then as a good approximation to $H_{b}$, and thus to $H_{\rm SB}$ whenever $|2\lambda/\Omega|\sqrt{\left<(a+\adag)^2\right>}\ll 1$ is not fulfilled, as recently shown in~\cite{Puebla:19}. In this article, however, we will constrain ourselves to parameters within the Lamb-Dicke regime.

\subsection{Structured environments}\label{ss:senv}

As previously mentioned, the simulation of multiphoton spin--boson interactions is not restricted to a determined form of $J_{\rm SB}(\omega)$. Here, we show the derivation of the procedure to obtain an effective multiphoton Hamiltonian when the initial spin--boson model features a more complicated interaction with the environment. For simplicity, we consider that $J_{\rm SB}(\omega)$ can be split in two parts, $J_{\rm SB}(\omega)=J_{\rm SB,1}(\omega)+J_{\rm SB,2}(\omega)$, although its generalization to more is straightforward. The first contribution, $J_{\rm SB,1}(\omega)$, is considered here to be suitable for the realization of multiphoton interactions as described in~\ref{ss:rc}. In addition, we will work under the assumption that the environment degrees of freedom corresponding to $J_{\rm SB,2}(\omega)$ can be treated and simplified using again a collective or reaction coordinate, as sketched in Figure~\ref{fig:scheme}{(c)}.

As discussed previously, we identify a collective coordinate for each of the contributions to the spectral density $J_{\rm SB}(\omega)$. In this manner, we augment the system to include both reaction coordinates, denoted here by ${\rm S'}={\rm{S}}+{\rm{RC}}_1+{\rm{RC}}_2$. Hence, its Hamiltonian is given by:
\begin{align}\label{eq:HSpr}
  H_{\rm S'}=H_{S,d}&+\Omega_1\adag_1 a_1+\lambda_1\sigma_x(a_1+\adag_1)\nonumber \\&+ \Omega_2 \adag_2 a_2+\lambda_2\sigma_x(a_2+\adag_2), 
\end{align}
 where $H_{\rm S,d}$ is the original spin Hamiltonian, which may contain spin rotations, introduced in Equation~(\ref{eq:sbmtime}), while the subscripts denote the corresponding reaction coordinate. The parameters $\lambda_i$ and $\Omega_i$ are determined by the spectral density $J_{\rm SB,i}(\omega)$. The dynamics of the augmented system $\rm S'$ is governed by the following master equation:
\begin{align}\label{eq:mesrc2}
\dot{\rho}_{\rm S'}(t)=&-i\left[H_{\rm S'},\rho_{\rm S'}(t)\right]\nonumber\\&- \left[x_1,\left[\chi_1,\rho_{\rm S'}(t) \right] \right] - \left[x_2,\left[\chi_2,\rho_{\rm S'}(t) \right] \right]\nonumber \\&+\left[x_1,\left\{\Theta_1,\rho_{\rm S'}(t) \right\} \right] + \left[x_2,\left\{\Theta_2,\rho_{\rm S'}(t) \right\} \right] ,
\end{align}
where $x_i = a_i + \adag_i$ for $i=1,2$, and $\chi_i$ and $\Theta_i$ are defined in analogy to Equations~(\ref{eq:chi})-(\ref{eq:theta}). 

In order to find a suitable transformation to realize multiphoton interaction terms from $H_{\rm S'}$, we proceed in a similar manner as for a single reaction coordinate. That is, we first move to a rotating frame where $H_{\rm S'} \equiv H_{a,1}^I$, with $H_{a}=H_{a,0}+H_{a,1}$ and $H_{a,0}=-\Delta_0/2\sigma_x$. Therefore, the transformed Hamiltonian reads as:
 \begin{align}
H_{a}= &\sum_{k=1,2} \Omega_k\adag_k a_k+\lambda_k\sigma_x(a_k+\adag_k)\nonumber\\+&\sum_j\frac{\epsilon_j}{2}\left[\cos\Delta_j t \sigma_z+\sin\Delta_j t\sigma_y \right].
 \end{align}
The next step is to perform the transformation using the unitary operator $T(\alpha)$. As previously mentioned, we consider that the first reaction coordinate is suitable for the quantum simulation of multiphoton interaction terms, due to the form of its spectral density. This argument enables one to choose $\alpha \equiv -\lambda_1/\Omega_1$, hence
$H_b\equiv T^{\dagger}(-\lambda_1/\Omega_1)H_a T(-\lambda_1/\Omega_1)$. This transformation acts trivially on the second reaction coordinate, but it does affect the coupling between the latter and the spin. Finally, if we move to an interaction picture with respect to $H_{b,0}=(\Omega_1-\tilde\nu_1)\adag_1a_1-\tilde\omega\sigma_z/2$, we obtain the Hamiltonian $H_{\rm n,2} \approx H_{b,1}^I\equiv U_{b,0}^{\dagger}H_{b,1}U_{b,0}$,
\begin{align}\label{eq:Hnsenv}
H_{\rm n,2} &=\frac{\tilde\omega}{2}\sigma_z+\tilde\nu\adag_1a_1 + \Omega_2 \adag_2 a_2- \lambda_2\sigma_z(a_2+\adag_2)\nonumber\\&+\sum_{j\in r}\frac{\epsilon_j}{2 n_j!} \left (\frac{2\lambda_1}{\Omega_1} \right)^{n_j}\left[\sigma^+a_1^{n_j}+{\rm H.c.}\right]\nonumber\\ &+ \sum_{j\in b}\frac{\epsilon_j}{2 n_j!} \left (\frac{2\lambda_1}{\Omega_1} \right)^{n_j} \left[\sigma^+(-\adag_1)^{n_j}+{\rm H.c.}\right],
\end{align}
where we have considered $\Delta_j=\pm n_j(\tilde\nu-\Omega_1)-\tilde\omega$ and assumed the Lamb-Dicke regime $|\lambda_1/\Omega_1|\sqrt{\langle(a_1+\adag_1)^2\rangle}\ll 1$, and $|\Omega_1-\tilde\nu|\gg \epsilon_j/2$ to perform a rotating-wave approximation. Note that, while the multiphoton terms are identical to those of $H_{\rm n}$ in Equation~(\ref{eq:Hn}), the second reaction coordinate interacts with the spin degree of freedom. Indeed, depending on the parameters of $H_{\rm n,2}$, the effect of such an interaction may turn effectively into non-Markovian effects for the reduced state of the spin and first reaction coordinate, $\rho_{\rm n}={\rm Tr}_2[\rho_{\rm n,2}]$. 
The final master equation governing the dynamics of $\rho_{\rm n,2}$ is:
\begin{align}\label{eq:rhonsenv}
\dot{\rho}_{\rm n,2}(t) &=-i[H_{\rm n,2},\rho_{\rm n,2}(t)]\nonumber\\ &- \left[\tilde{x}_1,\left[\tilde\chi_1,\rho_{\rm n,2}(t)\right]\right] - \left[\tilde{x}_2,\left[\tilde\chi_2,\rho_{\rm n,2}(t)\right]\right] \nonumber \\&+ \left[\tilde{x}_1,\left\{\tilde\Theta_1,\rho_{\rm n,2}(t)\right\}\right] + \left[\tilde{x}_2,\left\{\tilde\Theta_2,\rho_{\rm n,2}(t)\right \}\right]
\end{align}
where the operators involved are defined as in the case involving a single reaction coordinate (cf. Equation~(\ref{eq:merhon})). It is worth stressing that the relation between the states given in Equation~(\ref{eq:rhonSRC}) still holds. From the previous derivation, one can observe that the extension to more collective coordinates is straightforward.

\section{Examples and numerical simulations}\label{s:num}

In this section, we provide examples of the previously-explained general theoretical framework to investigate the performance of the quantum simulation of different multiphoton Hamiltonians $H_{\rm n}$, as well as to discuss the limitation in the parameter regime for their realization. In particular, in Section~\ref{ss:nondiss}, we first consider the case in which the original spin-boson model interacts just with a discrete number of modes, which can be viewed as a limit of vanishing spectral broadening $\Gamma\rightarrow 0$. This scenario will allow us to examine the validity of the required approximations without the effect of dissipation. Then, in Section~\ref{ss:diss}, we will consider $\Gamma\neq 0$, where the reaction-coordinate mapping appears as a key step to realize a desired multiphoton Jaynes-Cummings model. {The dynamics of each model is obtained by a standard numerical integration (fourth-order Runge-Kutta) of the corresponding master equation, namely Equations~(\ref{eq:mesrc}) and~(\ref{eq:merhon}) for the spin-boson and multiphoton Jaynes-Cummings model, respectively. Note that for a structured environment, the master equations are given in Equations~(\ref{eq:mesrc2}) and~(\ref{eq:rhonsenv}).}

In all cases, we assess the performance of the realization of the targeted multiphoton Jaynes-Cummings models by means of the fidelity $F(t)$ between two states,
\begin{align}\label{eq:Fid}
F(t)={\rm Tr}\left[\sqrt{\sqrt{\rho_{1}(t)}\rho_2(t)\sqrt{\rho_{1}(t)}}\right]^2.
\end{align}
In particular, we will analyse to what extent is the relation given in Equation~(\ref{eq:rhonSRC}) satisfied. {In other words, we will compare the aimed state of a multiphoton Jaynes-Cummings model $\rho_{\rm n}(t)$ with the one retrieved using the analogue simulator, $\Phi\rho_{\rm S+RC}(t)\Phi^{\dagger}$, that is $\rho_1(t)\rightarrow \rho_{\rm n}(t)$ and $\rho_{2}(t)\rightarrow \Phi\rho_{\rm S+RC}(t)\Phi^{\dagger}$ in Equation~(\ref{eq:Fid}).} We remark that when two reaction coordinates are included, the state $\rho_{\rm n}(t)$ obeys the master equation given in Equation~(\ref{eq:rhonsenv}), whose Hamiltonian is $H_{\rm n,2}$, Equation~(\ref{eq:Hnsenv}), while $\rho_{\rm S+RC}(t)$ must be replaced by $\rho_{\rm S'}$, as explained in~\ref{ss:senv}.

In addition, we will show that the theoretical framework allows us to realize non-Markovian multiphoton Jaynes-Cummings models. Among the different measures for non-Markovianity~\cite{deVega:17}, we resort to the one based on the trace distance~\cite{Breuer:09}, defined as: 
\begin{align}
\mathcal{D}(\rho_x,\rho_y)=\frac{1}{2}{\rm Tr}\left[\left|\rho_x-\rho_y \right|\right].
\end{align}
where $|A|=\sqrt{A^{\dagger}A}$. Then, non-Markovian evolutions can be characterized as those for which $\mathcal{D}(\rho_x(t),\rho_y(t))$ increases during certain time intervals, that is for those for which the time-derivative of the trace distance for a pair of states $\rho_{x,y}$, 
\begin{align}\label{eq:sigmaTD}
\sigma(t,\rho_{x,y})=\frac{d}{dt}\mathcal{D}(\rho_x(t),\rho_y(t)),
\end{align}
is $\sigma(t,\rho_{x,y})> 0$. In general, one has to maximize over all possible pairs of states $\rho_{x,y}$ in order to find a suitable non-Markovian measure~\cite{Breuer:09}. {For our purpose, however, it will be sufficient to show that $\sigma(t,\rho_{x,y})>0$ for a certain pair of states in a multiphoton Jaynes-Cummings model and that it can be retrieved using a spin-boson model. That is, we calculate $\sigma(t,\rho_{x,y})$ using two initial states $\rho_{x,y}$ in the multiphoton Jaynes-Cummings model and corroborate that $\sigma(t,\rho_{x,y})$ is obtained to a very good approximation when the states $\rho_{x,y}(t)$ are replaced by their simulated ones using the spin-boson model, namely $\rho_x(t)\rightarrow \Phi\rho_{\rm x,S+RC}(t)\Phi^{\dagger}$ and $\rho_y(t)\rightarrow \Phi\rho_{\rm y,S+RC}(t)\Phi^{\dagger}$.} In this manner, we offer a proof-of-principle that non-Markovian multiphoton models can be realized.

\subsection{Dissipationless multiphoton Jaynes-Cummings models}\label{ss:nondiss}
%
We start considering the simplest case, namely when the spin-boson model simply involves the interaction with a discrete number of modes. This corresponds to either considering $\Gamma\rightarrow0$ in the underdamped spectral density $J_{\rm SB}(\omega)$ or, equivalently, assuming that dissipation effects are sufficiently small so that they can be discarded. 
Note that for a single bosonic mode with $\Gamma=0$, the spin--boson model adopts the form of a generalized quantum Rabi model, which is indeed $H_{\rm S+RC}$, as given in Equation~(\ref{eq:HSRC}). Recall that in this particular case, $H_{\rm SB}\equiv H_{\rm S+RC}$, as there are no further modes in the system. In particular, we set $n_d=0$ in Equation~(\ref{eq:sbmtime}) as we aim to realize a single multiphoton Jaynes-Cummings interaction. The Hamiltonian for a nJCM can be written in general as:
\begin{align}
H_{\rm nJCM}=\frac{\tilde\omega}{2}\sigma_z+\tilde\nu\adaga+\tilde{g}_n\left(\sigma^+a^n+\sigma^-(\adag)^n\right).
\end{align}
At resonant condition, $\tilde\omega=n\tilde\nu$, the coupling constant $\tilde{g}_n$ fixes the time required to transfer the population from the state $\ket{e}\ket{0}$ to $\ket{g}\ket{n}$, denoted as $\tau_n=\pi/(2\tilde{g}_n\sqrt{n!})$. Both are related to the spin-boson parameters as (cf. Equation~(\ref{eq:Hn})):
\begin{align}
 \tilde{g}_n&=\frac{\epsilon_0}{2\ n!}\left(\frac{2\lambda}{\Omega}\right)^n\\ \label{eq:taun}
 \tau_n&=\frac{\sqrt{n!}}{\epsilon_0}\left(\frac{\Omega}{2\lambda}\right)^n. 
\end{align}
Clearly, as $2\lambda/\Omega$ must be small to lie within the Lamb-Dicke regime, the coupling $\tilde{g}_n$ decreases considerably for increasing $n$, requiring longer evolution times under the spin-boson Hamiltonian to observe a significant effect, that is an evolution time of the order of $\tau_n$.

\begin{figure}
\centering
\includegraphics[width=0.83\linewidth,angle=-90]{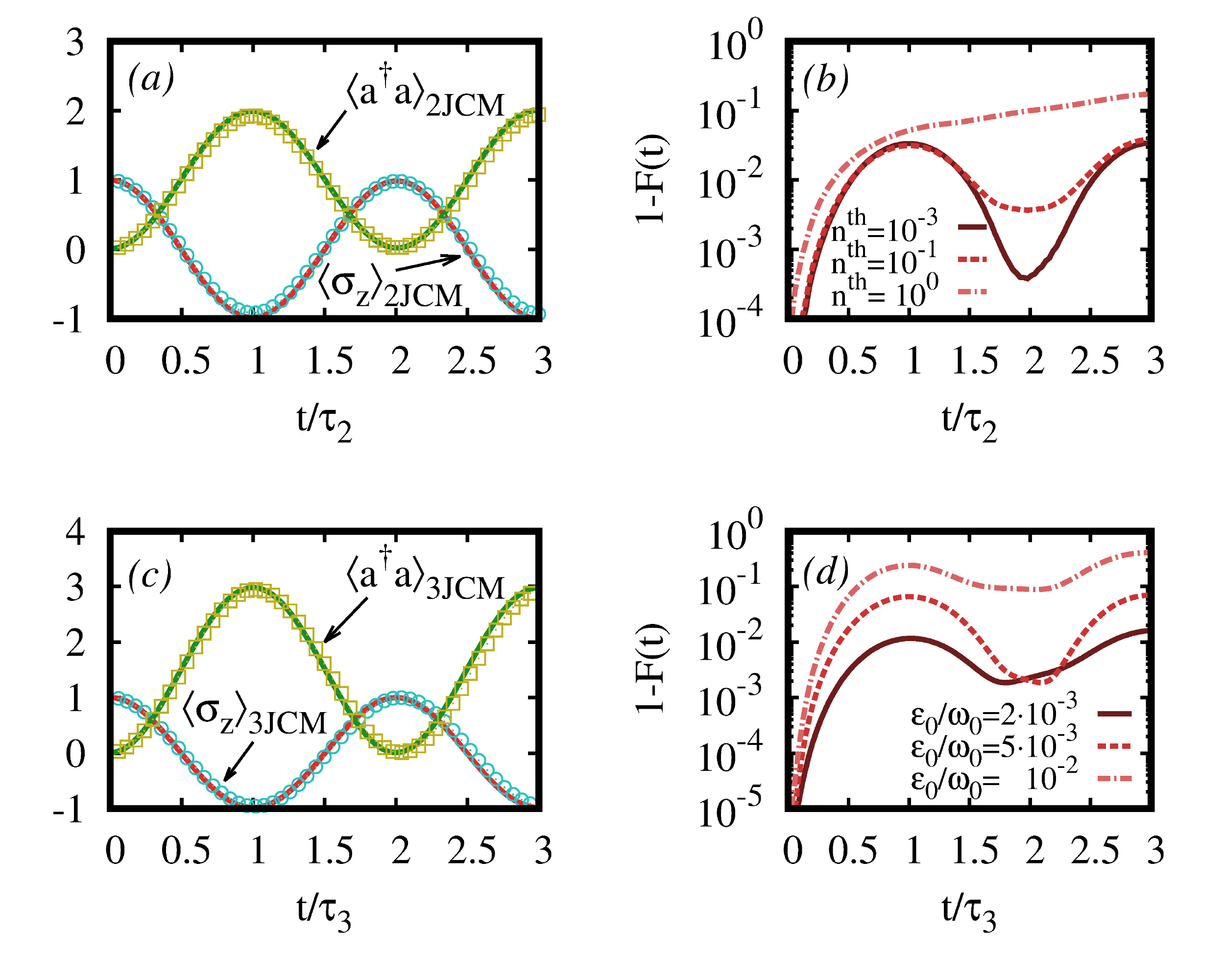}
\caption{\small{{Dynamics of the simulated multiphoton }
 Jaynes-Cummings models, $n=2$ (top) and $n=3$ (bottom). In Panels {(a)} and {(c)}, we show the targeted dynamics (solid lines) and the one obtained using the spin-boson Hamiltonian (points) for $\left<\adaga\right>$ and $\left<\sigma_z\right>$, as indicated in the plots and as a function of the time rescaled by $\tau_n$ (Equation~(\ref{eq:taun})). In Panels {(b)} and {(d)}, we plot the infidelity $1-F(t)$ between the ideal $\rho_{\rm nJCM}(t)$ state and its approximated one $\Phi \rho_{\rm S+RC}(t)\Phi^{\dagger}$ for different conditions, namely in {(b)} for different temperatures (or mean occupation number $n^{\rm th}$) and in {(d)} for different values of $\epsilon_0/\Omega$. See~\ref{ss:nondiss} for further details regarding the parameters and states considered in the simulation. JCM, Jaynes-Cummings model.}}
\label{fig2}
\end{figure}

In Figure~\ref{fig2}, we show the results for the realization of 2JCM and 3JCM models using a spin-boson model interacting with a single bosonic mode. In order to observe the paradigmatic Rabi oscillations between the states $\ket{e}\ket{0}$ and $\ket{g}\ket{n}$, we choose $\rho_{\rm S+RC}(0)=\ket{-}\bra{-}\otimes \rho_{\rm RC}^{\rm th}$ as an initial state for the spin-boson model, where $\rho_{\rm RC}^{\rm th}$ is a thermal state at temperature $\beta^{-1}$ for the reaction coordinate mode, containing $n^{\rm th}=(e^{\beta\Omega}-1)^{-1}$ bosons. Recall that, as we consider here a single spectral density with $\Gamma=0$, the reaction coordinate mode is simply the unique mode that interacts with the spin degree of freedom. In this manner, the initial state for the simulated multiphoton models reads as $\rho_{\rm nJCM}(0)=T^{\dagger} \rho_{\rm S+RC}(0)T$, which approximately amounts to $\rho_{\rm nJCM}(0)\approx \ket{e}\bra{e}\otimes \ket{0}\bra{0}$ for sufficiently low temperature and small $2\lambda/\Omega$. 
The chosen parameters for the simulation of the 2JCM, plotted in Figure~\ref{fig2}{(a)} and {(b)}, are $\pi\alpha=\epsilon_0=0.02\omega_0$; recalling that $\Omega=\omega_0$, it results in $2\lambda/\Omega=0.2$. Choosing $\tilde\nu=10^{-3}\Omega$ and $\tilde\omega=2\tilde\nu$, the coupling in 2JCM amounts to $\tilde{g}_2=0.2\tilde\nu$. The initial reaction-coordinate thermal state, $\rho_{\rm RC}^{\rm th}$, contains $n^{\rm th}=10^{-3}$ bosons. In Figure~\ref{fig2}{(b)}, we show how the quantum simulation of the 2JCM model deteriorates for increasing number of bosons, as a large $n^{\rm th}$ will eventually break down the Lamb-Dicke regime.

For the 3JCM, we choose again $\pi\alpha=0.02\omega_0$, which leads to $2\lambda/\Omega=0.2$. Then, we select the aimed coupling strength of the multiphoton interaction to be $\tilde{g}_3=0.1\tilde\nu$ with $\tilde\omega=3\tilde\nu$, while we vary $\epsilon_0/\omega_0$. The temperature is set to $\beta\Omega\approx 100$ so that $\rho_{\rm RC}^{\rm th}\approx \ket{0}\bra{0}$. As in the previous case, the dynamics are well retrieved; see Figure~\ref{fig2}{(c)}, where we have set $\epsilon_0/\omega_0=2\cdot 10^{-3}$. Note however that, as a consequence of the rotating-wave approximation performed to achieve a resonant third order (see Appendix~\ref{app:b} and cf. Equation~(\ref{eq:Hn})) and due to the longer times required to simulate a 3JCM compared to the 2JCM, the condition $|\Omega-\tilde\nu|\gg \epsilon_0$ must be better satisfied. Indeed, for $\epsilon_0/\omega_0=10^{-2}$, we already see a clear departure from the targeted dynamics, as indicated by a large infidelity $1-F(t)\gtrsim 10^{-1}$, as shown in Figure~\ref{fig2}{(d)}.

\begin{figure}
\centering
\includegraphics[width=0.83\linewidth,angle=-90]{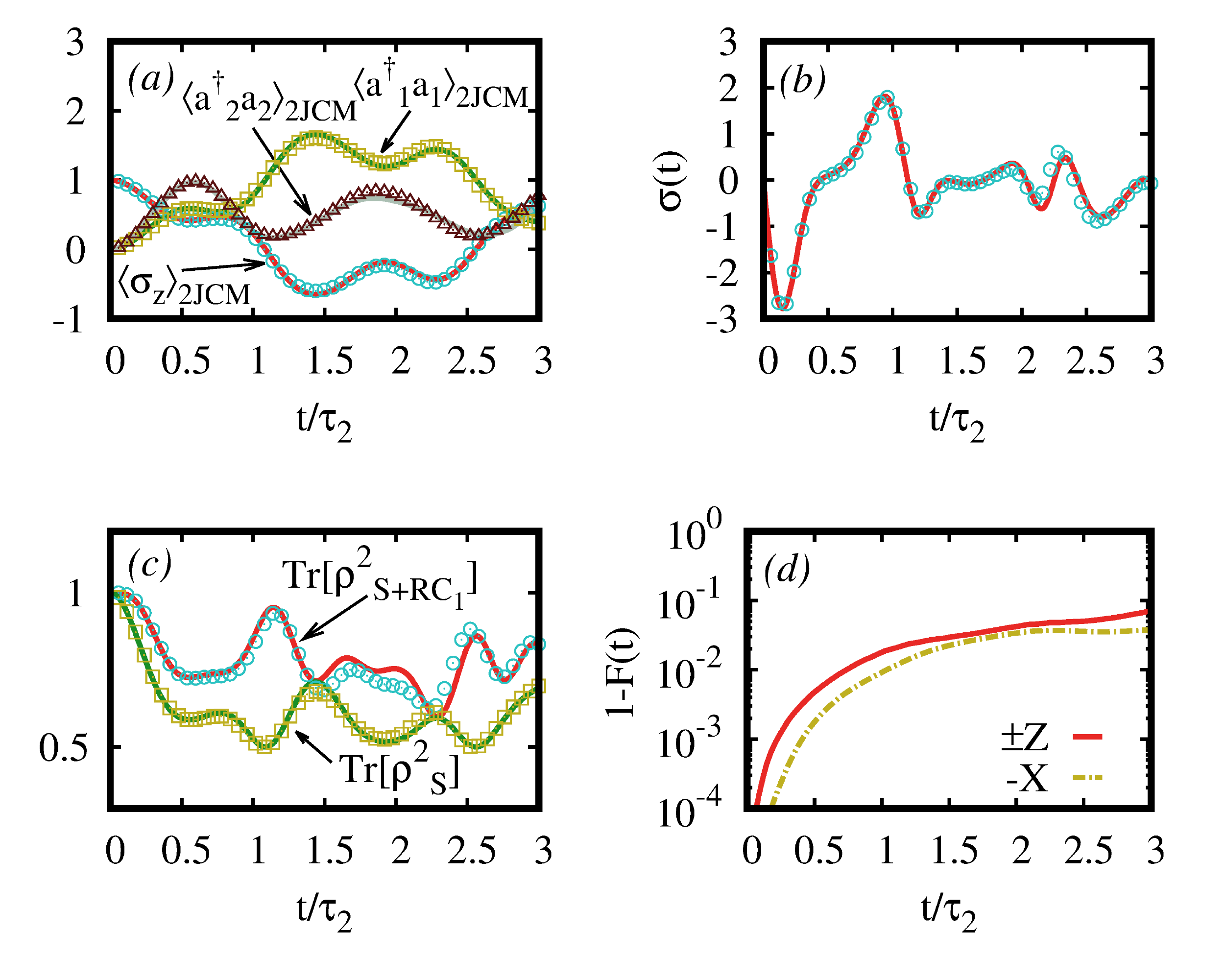}
\caption{\small{{Non-Markovian dynamics for a 2JCM and its simulation} 
using a spin-boson model $H_{\rm S'}$. In Panel {(a)}, we show the dynamics for the expectation values $\langle a_i^{\dagger}a_i\rangle$ with $i=1,2$ and $\langle\sigma_z\rangle$ for the target 2JCM model (solid lines) and its reconstructed values using $H_{\rm S'}$ (points). The considered initial state reads as $\rho_{\rm S'}(0)=\ket{-}\bra{-}\otimes\rho_{\rm RC_1}^{\rm th}\otimes\rho_{\rm RC_2}^{\rm th}$, with $\beta$ very large such that $\rho^{\rm th}\approx \ket{0}\bra{0}$. In {(b)}, we plot the time-derivative of the trace distance, $\sigma(t)$, after tracing out the second bosonic mode and considering the initial states $\ket{e}$ and $\ket{g}$ for the spin in $H_{\rm S'}$, while both reaction coordinates find themselves in their vacuum. Clearly, $\sigma(t)>0$ during certain intervals, revealing the non-Markovianity introduced due to the interaction with the second mode. Panel {(c)} shows the evolution of purity for the state upon tracing the second mode, ${\rm Tr}[\rho_{\rm S+RC_1}^2(t)]$ and for the reduced state of the spin, ${\rm Tr}[\rho_{\rm S}^{2}(t)]$, for the same case shown in {(a)}. In Panel {(d)}, we compare the infidelity $1-F(t)$ between the ideal state and the simulated one using $H_{\rm S'}$ for the three different initial states employed here. We refer to Section~\ref{ss:nondiss} for further details regarding the parameters and states considered in the simulation. }}
\label{fig3}
\end{figure}

In the following, we consider a spin interacting with two bosonic modes, again with $\Gamma_{1,2}=0$. As explained in~\ref{ss:senv}, we perform the map onto the first bosonic mode to attain a multiphoton interaction. Upon suitable transformations and approximations, the spin-boson model will take the form of a multiphoton Jaynes-Cummings model $H_{\rm nJCM,2}$, where the subscript $2$ indicates the presence of a second reaction coordinate in the system. The Hamiltonian $H_{\rm nJCM,2}$ reads as:
\begin{align}
H_{\rm nJCM,2}&=\frac{\tilde\omega}{2}\sigma_z+\tilde\nu\adag_1a_1+\Omega_2\adag_2a_2\nonumber\\&+\tilde{g}_n\left(\sigma^+a_1^n+\sigma^-(\adag_1)^n\right)-\lambda_2\sigma_z(a_2+\adag_2).
\end{align}
In this manner, the spin exchanges $n$ quanta with the first bosonic mode as in $H_{\rm nJCM}$, while the last term effectively shifts the spin frequency depending on the state of the second mode. The reduced state for the spin and first bosonic mode is given then by $\rho_{\rm nJCM}(t)={\rm Tr}_{2}[\rho_{\rm nJCM,2}(t)]$. Indeed, due to the interaction with the second bosonic mode, the multiphoton Jaynes-Cummings model may exhibit non-Markovian features.
For that, we consider the spin-boson Hamiltonian $H_{\rm S'}$ given in Equation~(\ref{eq:HSpr}), which then approximately realizes $H_{\rm nJCM,2}$. In particular, we select $\Delta_0=-2\Omega_1$, so that the simulated model involves two-photon interaction terms, i.e., a 2JCM. The results are plotted in Figure~\ref{fig3}, while the parameters are $\pi\alpha_i=0.02\Omega_i$ such that $2\lambda_i/\Omega_i=0.2$ for $i=1,2$, $\epsilon_0/\Omega_1=10^{-2}$. The coupling strength in $H_{\rm 2JCM,2}$ is given by $\tilde{g}_2=0.2\tilde\nu$ with $\tilde\nu=\Omega_2$. As in the single-mode case, Rabi oscillations will be clearly visible selecting $\rho_{\rm S'}(0)=\ket{-}\bra{-}\otimes \rho_{\rm RC_1}^{\rm th}\otimes \rho_{\rm RC_2}^{\rm th}$. After its transformation, this state corresponds approximately to an initial spin state $\ket{e}$ in the nJCM frame. In the same manner, in order to analyse the emergence {of} non-Markovian behaviour, we consider the initial states $\ket{g}\bra{g}$ and $\ket{e}\bra{e}$ for the spin in $H_{\rm S'}$. This implies initial spin states $\ket{\pm}$ in the nJCM frame, which for pure dephasing noise, it has been shown to be the pair of states maximizing $\sigma(t)$~\cite{Breuer:09}. The results plotted in Figure~\ref{fig3} have been performed considering a sufficiently low temperature such that $\rho^{\rm th}_{\rm RC_{1,2}}\approx \ket{0}\bra{0}$. We then compute the trace distance $\mathcal{D}(\rho_x,\rho_y)$ using the states $\rho_{x,y}(t)$ resulting in tracing out the second mode, ${\rm Tr}_{\rm 2}[\rho_{\rm 2JCM,2}(t)]$, for the two different initial states $\rho_{2JCM,2}(0)\approx \ket{\pm}\bra{\pm}\otimes \rho_{\rm RC_{1}}^{\rm th}\otimes \rho_{\rm RC_{2}}^{\rm th}$. 
As shown in Figure~\ref{fig3}{(b)}, the time-derivative of the trace distance, $\sigma(t)$, becomes positive during certain intervals, a clear indication of the non-Markovian behaviour of the simulated multiphoton Jaynes-Cummings model. In addition, we also calculate the non-trivial evolution of the purity for the states $\rho_{\rm S+RC_1}(t)$ and $\rho_{\rm S}(t)={\rm Tr}_{\rm RC_1}[\rho_{\rm S+RC_1}(t)]$, which is shown in Figure~\ref{fig3}{(c)}. According to our theoretical framework, their purity is approximately equal to that of $\rho_{\rm 2JCM}(t)$ and the reduced spin state upon tracing both bosonic degree of freedom in the 2JCM, ${\rm Tr}_{}[\rho_{\rm 2JCM}(t)]$, respectively. Finally, the infidelity $1-F(t)$ between the targeted state $\rho_{\rm 2JCM,2}(t)$ and its reconstructed one $\Phi \rho_{\rm S+RC_1+RC_2}(t)\Phi^{\dagger}$ in Figure~\ref{fig3}{(d)}.

\subsection{Dissipative multiphoton Jaynes-Cummings models}\label{ss:diss}

\begin{figure}
\centering
\includegraphics[width=0.83\linewidth,angle=-90]{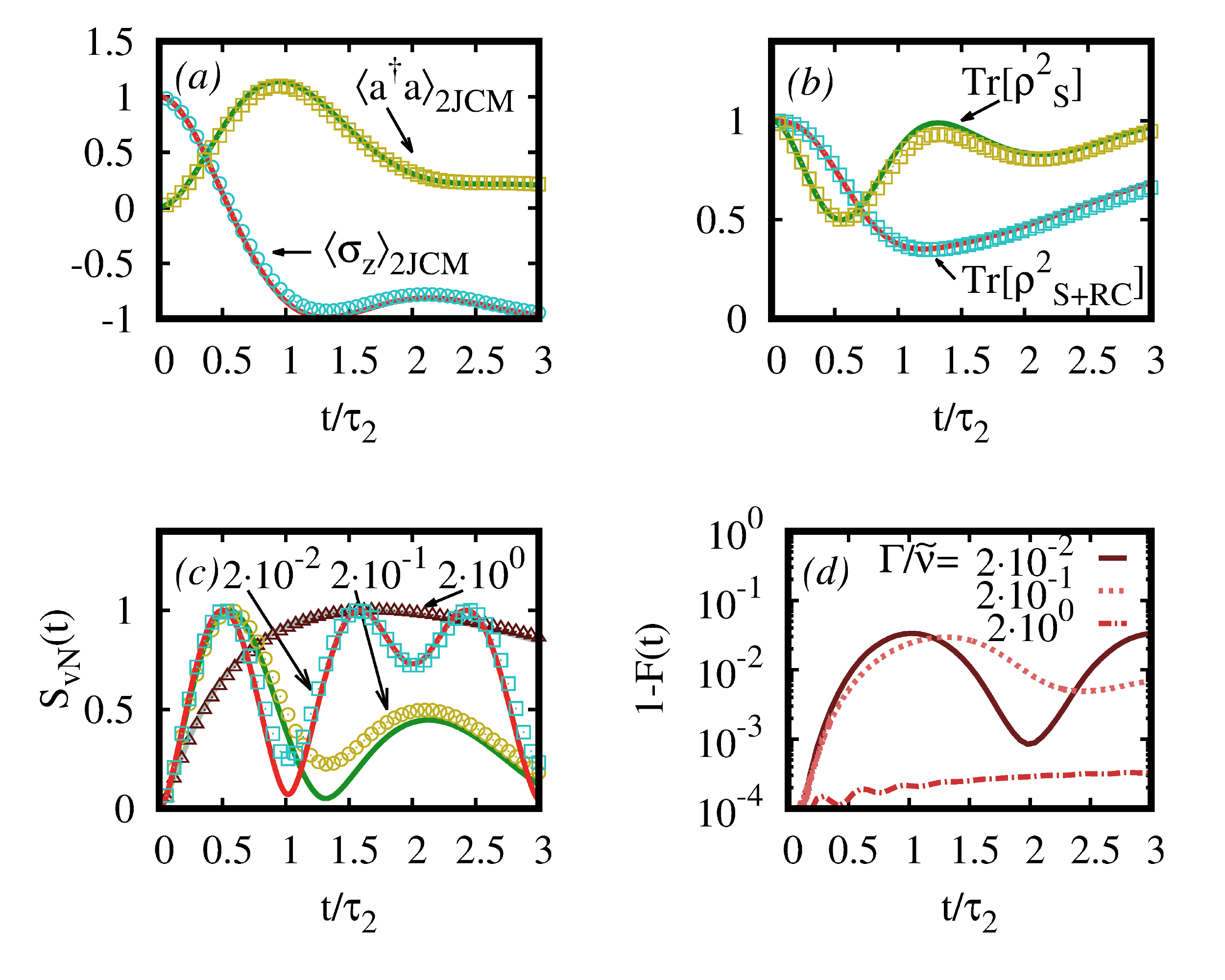}
\caption{\small{{Dynamics of a dissipative 2JCM using a spin-boson model.} 
In Panel {(a)}, we show the dynamics of the expectation values of $\left<\adaga\right>$ and $\left<\sigma_z\right>$, as in Figure~\ref{fig2}, for the dissipative 2JCM (solid lines) and its simulation using the spin-boson model (points), for $\Gamma/\tilde\nu=2\cdot10^{-1}$ and $\rho_{\rm S+RC}(0)=\ket{-}\bra{-}\otimes\rho_{\rm RC}^{\rm th}$ with $n^{\rm th}=10^{-3}$. For the same case, we also show in {(b)} the evolution of the purities for the spin state ${\rm Tr}[\rho_{\rm S}^2(t)]$ and for the total state ${\rm Tr}[\rho_{\rm S+RC}^2(t)]$. In {(c)}, we compare the different behaviour as $\Gamma/\tilde\nu$ varies for the von Neumann entropy of the reduced spin state, $S_{\rm vN}(\rho_{\rm S}(t))$. The values of $\Gamma/\tilde\nu$ are indicated close to each curve. Finally, the state infidelity $1-F(t)$ between the targeted $\rho_{\rm 2JCM}$ and its approximate simulation, $\Phi\rho_{\rm S+RC}(t)\Phi^\dagger$, is plotted in Panel {(d)} for different $\Gamma/\tilde\nu$. See the main text for further details on the parameters employed for the simulation. }}
\label{fig4}
\end{figure}

We now consider a more realistic scenario in which the spin-boson model interacts with an environment whose spectral density has an underdamped shape, i.e., $J_{\rm SB}(\omega)$ has the form of Equation~(\ref{eq:JSBUD}) with $\Gamma\neq 0$. In this manner, we extend the theoretical framework beyond the standard local master equation description~\cite{Puebla:19}. As explained in Section~\ref{ss:rc}, this situation can be mapped using a reaction coordinate, which now in turn interacts with a Markovian residual environment. The evolution of the state of the augmented system, spin and reaction coordinate, evolves according to the master equation given in~(\ref{eq:mesrc}). Indeed, the effect of spectral broadening, $\Gamma\neq 0$, introduces dissipation into the simulated multiphoton Jaynes-Cummings model, whose state now obeys the master equation~(\ref{eq:merhon}). We remark that the performance of the simulated dissipative model is not altered when the effect of dissipation is taken into account correctly. Nevertheless, whenever $\Gamma\gg \tilde\nu$, dissipation dominates the dynamics, and the paradigmatic Rabi oscillations will eventually fade away.
In Figure~\ref{fig4}, we show the results of numerical simulations aimed to retrieve a 2JCM with different $\Gamma/\tilde\nu$ values and for different quantities. As for Figure~\ref{fig2}, we used $\pi\alpha=\epsilon_0=0.02\omega_0$, so that $2\lambda/\Omega=0.2$. We chose again $\tilde\nu=10^{-3}\Omega$ and $\tilde\omega=2\tilde\nu$, and therefore, the coupling in 2JCM amounts to $\tilde{g}_2=0.2\tilde\nu$, while the temperature is such that $\rho_{\rm RC}^{\rm th}$ contains $n^{\rm th}=10^{-3}$ bosons. The spin is initialized in the $\ket{-}$ state, so that $\rho_{\rm S+RC}(0)=\ket{-}\bra{-}\otimes\rho_{\rm RC}^{\rm th}$. In particular, the value $\Gamma/\tilde{\nu}=2\cdot 10^{-1}$ considered in Figure~\ref{fig4}{(a)} already produces a significant departure from the Rabi oscillation between the states $\ket{e}\ket{0}$ and $\ket{g}\ket{2}$ in the dissipationless 2JCM (cf. Figure~\ref{fig2}{(a)} for $\Gamma=0$). Note that the results plotted in Figure~\ref{fig4}{(a)} correspond to a critically-damped 2JCM since $\Gamma=\tilde{g}_2$. As plotted in Figure~\ref{fig4}{(b)}, the effect of the dissipation is clearly visible in the evolution of the purity for both the total state (spin plus bosonic mode) and the reduced spin state, namely ${\rm Tr}[\rho_{\rm S+RC}^2(t)]$ and ${\rm Tr}[\rho_{\rm S}^2(t)]$. As in previous cases, the purity of these states is directly related to those of the simulated model as a consequence of the relation $\rho_{\rm 2JCM}(t)\approx \Phi \rho_{S+RC}(t)\Phi^\dagger$. Furthermore, Rabi oscillations or population revivals appear in the evolution of von Neumann entropy, $S_{\rm vN}(\rho)=-\rho\log_2\rho$ for the reduced spin state. {In particular, for an initial state $\rho_{nJCM}(0)\approx \ket{e}\bra{e}\otimes \ket{0}\bra{0}$ and due to the $n$-photon interaction with a bosonic degree of freedom, the spin state oscillates between a pure ($S_{\rm vN}=0$) and a maximally-mixed state ($S_{\rm vN}=1$) in a time $\tau_n/2$. This further corroborates that one can witness the multiphoton transitions of the aimed multiphoton Jaynes-Cummings model monitoring the spin even without access or control on the bosonic environment.}
This is plotted in Figure~\ref{fig4}{(c)} for different $\Gamma/\tilde\nu$ values. Finally, we note that the performance of the quantum simulation is independent of the dissipation as demonstrated by the good fidelities attained in these cases (cf. Figure~\ref{fig4}{(d)}), allowing for the simulation of different parameter regimes in a nJCM. 

\section{Conclusions}\label{s:conc}
We have proposed a theoretical scheme to realize multiphoton Jaynes-Cummings models using the paradigmatic spin-boson model, which contains a continuum of bosonic modes, as an analogue quantum simulator. While the spin-boson model naturally lacks these multiphoton interaction terms, we make use of a suitable transformation that approximately maps the spin--boson model into a dissipative multiphoton Jaynes-Cummings model. {Importantly, the parameters of the multiphoton model, as well as the order of the interaction can be controlled by tuning the frequency splitting and bias parameter of the spin in the original spin-boson model.} In order to bring the spin-boson model, typically interacting with an infinite number of bosonic modes, into the form of the aimed multiphoton model, we first rearrange the environment degrees of freedom using the so-called reaction-coordinate method~\cite{Thoss:01,Martinazzo:11,Iles:14,Iles:16,Strasberg:16,Strasberg:18,Nazir:18}. This method allows us to include a set of collective bosonic modes into the coherent description of the problem, which then in turn interact with the residual environment. For certain types of interactions between the spin and the environment, characterized by the spectral density, the reaction coordinate mapping emerges as a powerful tool to reduce the complexity of the problem. In particular, for an underdamped spectral density, the reaction coordinate takes a simple form as it interacts with the residual environment in a Markovian fashion. The resulting Hamiltonian is then used to generate multiphoton interaction terms, following the theory explained in~\cite{Casanova:18,Puebla:19}, while the dissipation effects must be transformed accordingly. Furthermore, we extend the scheme to spin-boson models with structured environments. In these cases, the original spin-boson Hamiltonian can be mapped onto the one of a spin interacting with more reaction coordinates. In this manner, we show how to extend the theoretical framework to account for these additional modes. In particular, due to the presence of two or more reaction coordinates, the attained multiphoton Jaynes-Cummings model can exhibit non-Markovian features. 
We perform numerical simulations starting from the spin plus reaction-coordinate Hamiltonians and aiming to realize different multiphoton Jaynes-Cummings models. We first perform simulations considering one reaction coordinate without dissipation to better illustrate the performance of the required approximations to achieve two- and three-photon Jaynes-Cummings models. We then demonstrate that non-Markovian multiphoton Jaynes-Cummings models can be indeed attained when a second reaction coordinate is included, as unveiled by the standard trace distance measure~\cite{Breuer:09}. Finally, we provide numerical simulations investigating the interplay between spectral broadening, dissipation and the decoherence in the targeted multiphoton models.

\acknowledgments
G.Z. is supported by the H2020-MSCA
-COFUND
-2016 project SPARK
 (Grant No. 754507). R.P. and M.P. acknowledge the support by the SFI-DfE
 Investigator Programme (Grant 15/IA/2864). M.P. acknowledges the H2020 Collaborative Project TEQ
 (Grant Agreement 766900), the Leverhulme Trust Research Project Grant UltraQuTe (Grant No. RGP-2018-266) and the Royal Society Wolfson Fellowship (RSWF\textbackslash R3\textbackslash183013). J.C. acknowledges support by the Juan de la Cierva Grant IJCI-2016-29681. I.A. acknowledges support by Basque Government Ph.D. Grant No. PRE-2015-1-0394. We also acknowledge funding from Spanish MINECO/FEDER FIS2015-69983-P and Basque Government IT986-16. This material is also based on work supported by the U.S. Department of Energy, Office of Science, Office of Advance Scientific Computing Research (ASCR), Quantum Algorithm Teams (QAT) Program under Field Work Proposal Number ERKJ333. J.C. and E.S. acknowledge support from the projects QMiCS
 (820505) and OpenSuperQ
 (820363) of the EU Flagship on Quantum Technologies.

\appendix

\section{Reaction coordinate mapping}\label{app:a}

In this Appendix, we provide the necessary steps for the reaction coordinate mapping, as well as for the derivation of the master equation given in Equation~(\ref{eq:mesrc}), following closely~\cite{Iles:14}.
As outlined in Section \ref{ss:rc}, given the Hamiltonian of the spin--boson system $H_ {\rm SB} = \frac{\epsilon_0}{2}\sigma_z+\frac{\Delta_0}{2}\sigma_x+\sigma_x\sum_k f_k (c_k+c_k^{\dagger})+\sum_k \omega_k c_k^{\dagger}c_k$, one can achieve the RC mapping by defining a collective coordinate such that $\lambda(a+\adag)=\sum_k f_k(c_k+c^{\dagger}_k)$, where $a$ and $\adag$ are respectively the annihilation and creation operators of the RC.
This transformation leads to a new Hamiltonian where the original system interacts with the residual environment only through the RC,
\begin{align}
H = H_{\rm S+RC} + H_{\rm RC-E'} + H_{\rm E'},
\end{align}
where $H_{\rm S+RC}$ is given by Equation (\ref{eq:HSRC}), while $H_{\rm RC-E'}=(a+\adag)\sum_k g_k(b_k+b_k^{\dagger})+(a+\adag)^2\sum_k\frac{g_k^2}{\omega_k}$, $H_{\rm E'} = \sum_k \omega_kb_k^{\dagger}b_k$.

The crucial point of this procedure is to find an explicit relation between the spectral density of the original configuration, i.e. $J_{\rm SB}(\omega)=\sum_k f_k^2\delta(\omega-\omega_k)$, and the analogue quantity of the transformed system $J_{\rm RC}(\omega)=\sum_k g_k^2\delta(\omega-\omega_k)$. In order to obtain this relation, one can rephrase the problem classically. Indeed, since the spectral density only depends on the interaction between the system and the environment, one can momentarily regard the spin as a continuous coordinate $q$ subject to a potential $V(q)$. After solving the corresponding Hamilton equations of motion in the Fourier space, one obtains an equation of the form
$\hat{L}_{\rm SB}(z) \hat{q} (z) = - \hat{V}'(z)$,
where $\hat{L}_{\rm SB}(z) = -z^2\left ( 1 + \int_{0}^{+\mathcal{1}} d \omega \frac{2 J_{\rm SB}(\omega)}{\omega (\omega^2 - z^2)}\right )$. Therefore, using the so-called Leggett prescription, one gets: 
\begin{equation}
\label{eq:JSB-rc}
 J_{\rm SB}(\omega) = \frac{1}{\pi} \lim_{\epsilon \to 0^{+}} {\rm{Im}}\left[ {\hat{L}}_{{\rm SB}} (\omega - i \epsilon) \right].
\end{equation}
One can reproduce the same calculation also after performing the RC mapping and express $J_{\rm RC}(\omega)$ in terms of the corresponding kernel $\hat{L}_0(z)$.
However, since at this stage, we are just rearranging the environment in a more convenient way by using a suitable normal mode transformation, the integral kernel must be the same before and after the mapping; hence, one can use $\hat{L}_0(z)$ instead of $\hat{L}_{\rm SB}(z)$ in Equation (\ref{eq:JSB-rc}).
By considering the Ohmic spectral density $J_{\rm RC}(\omega)=\gamma\omega e^{-\omega/\Lambda}$, one obtains:
\begin{align}
J_{\rm SB}(\omega) = \frac{4 \gamma \Omega^2 \lambda^2 \omega}{(\Omega^2 - \omega^2)^2 + (2 \pi \gamma \Omega \omega)^2}.
\end{align}
It is easy to see that one exactly recovers the underdamped spectral density given by Equation (\ref{eq:JSBUD}) by simply requiring that $\gamma=\Gamma/(2\pi\omega_0)$, $\Omega=\omega_0$, and $\lambda=\sqrt{\pi\alpha\omega_0/2}$. Furthermore, one also needs to solve the dynamics, i.e., writing down the corresponding master equation for the mapped system, system plus reaction coordinate. The guiding idea is to treat exactly the coupling between the spin and the RC, while the interaction between the latter and the residual environment is treated perturbatively up to the second order. This enables us to rely on the standard Born--Markov approximation, provided that either the coupling between the augmented system and the residual environment is weak or the residual environment correlation time is short compared to the relevant time scale of the system. Within this approximation, one can work out a master equation that, in the Schr\"{o}dinger picture, reads as:
\begin{align}
\label{eq:me-bm}
&\dot{\rho}(t)=-i\left[H_{\rm S+RC},\rho(t)\right]\\ &- \int\limits_{0}^{\mathcal{1}} d \tau \int\limits_{0}^{\mathcal{1}} d \omega J_{\rm RC}(\omega) \cos{\omega \tau} \coth \left(\frac{\beta \omega}{2} \right) \left[ A, \left[A(-\tau), \rho (t) \right] \right]\nonumber\\
&- \int\limits_{0}^{\mathcal{1}} d \tau \int\limits_{0}^{\mathcal{1}} d \omega J_{\rm RC}(\omega) \frac{\cos{\omega \tau}}{\omega} \left[ A, \left \{ \left[A(-\tau), H_{\rm S+RC} \right],\rho(t) \right \} \right],\nonumber
\end{align}
where $\rho \equiv \rho_{\rm S+RC}$, $A = a + \adag$, and the residual environment is assumed to be in a thermal state, i.e., $\rho_{E'} = e^{- \beta H_{\rm E'}} / {\rm{Tr}}_{\rm E'} \{ e^{- \beta H_{\rm E'}}\}$.

In order to obtain an expression for the interaction picture operators, one can proceed by truncating the space of the augmented system up to $n$ basis states and numerically diagonalising the Hamiltonian $H_{\rm S+RC}$. To this end, let $\ket{\phi_n}$ be an eigenstate of $H_{\rm S+RC}$, i.e., $H_{\rm S+RC} \ket{\phi_j} = \varphi_j \ket{\phi_j}$; therefore, the operator $A$ can be expanded as 
$A = \sum_{jk} A_{jk} \ket{\phi_j} \bra{\phi_k}$,
while in the interaction picture, one has:
\begin{align}
\label{eq:Aexp}
A(t) = \sum_{jk} A_{jk} e^{i \xi_{jk}t} \ket{\phi_j} \bra{\phi_k},
\end{align}
where $A_{jk} = \bra{\phi_j} A \ket{\phi_k}$, and $\xi_{jk} = \varphi_j -\varphi_k$. Finally, by plugging Equation (\ref{eq:Aexp}) into Equation (\ref{eq:me-bm}) and assuming the imaginary parts to be negligible, one gets the final form of the master equation given by Equation (\ref{eq:mesrc}).

\section{Derivation of $H_{b}$ and $H_{\rm n}$}\label{app:b}
In this Appendix, we show how to obtain the Hamiltonians $H_b$ and $H_{\rm n}$, given in Equations~(\ref{eq:Hb}) and~(\ref{eq:Hn}), respectively. In particular, for $H_b$, the following expressions are needed:
\begin{align*}
 T^{\dagger}(\alpha)\adaga T(\alpha)&=\adaga+|\alpha|^2-\sigma_z(a \alpha^*+\adag\alpha),\\
 T^{\dagger}(\alpha)\sigma_xT(\alpha)&= - \sigma_z,\\
 T^{\dagger}(\alpha)\sigma_y T(\alpha)&= -iD(2\alpha)\sigma^++{\rm H.c.},\\
 T^{\dagger}(\alpha)\sigma_z T(\alpha)&= D(2\alpha)\sigma^++{\rm H.c.},\\
 T^{\dagger}(\alpha)\sigma_x(a+\adag) T(\alpha)&=-\sigma_z(a+\adag)+2{\rm Re}[\alpha].
\end{align*}
Thus, the resulting Hamiltonian $H_b=T^{\dagger}H_{a}T$, with $H_a=\Omega\adaga+\lambda\sigma_x(a+\adag)+\sum_j\epsilon_j/2(\cos\Delta_jt\sigma_z+\sin\Delta_jt\sigma_y)$, reads:
\begin{align}
H_b=
&=\Omega\adaga-\Omega\sigma_z(a\alpha+\adag \alpha^*)-\lambda\sigma_z(a+\adag)\nonumber\\&+\sum_{j=0}^{n_d}\frac{\epsilon_j}{2}\left[ \sigma^+D(2\alpha)e^{-i\Delta_j t}+{\rm H.c.}\right],
\end{align}
where we have neglected a constant energy shift. Therefore, by selecting $\alpha=-\lambda/\Omega$, we obtain a simple Hamiltonian to pursue multiphoton interactions, namely:
\begin{align}
H_b=\Omega\adaga+\sum_j\frac{\epsilon_j}{2}\left[\sigma^+e^{2\lambda(a-\adag)/\Omega}e^{-i\Delta_j t}+{\rm H.c.} \right],
\end{align}
which is indeed Equation~(\ref{eq:Hb}). Moving now to an interaction picture w.r.t. $H_{b,0}=(\Omega-\tilde\nu)\adaga-\tilde\omega\sigma_z/2$, we obtain:
\begin{equation}
\begin{aligned}
H_{b,1}^I&
=\tilde\nu\adaga+\frac{\tilde\omega}{2}\sigma_z\\&+\sum_j\frac{\epsilon_j}{2}\left[\sigma^+e^{-i(\Delta_j+\tilde\omega)t}e^{2\lambda(a(t)-\adag(t))/\Omega}+{\rm H.c.} \right]
 \end{aligned}
\end{equation}
with $a(t)=ae^{-i(\Omega-\tilde\nu)t}$. Requiring $|2\lambda/\Omega|\sqrt{\left<(a+\adag)^2\right>}\ll 1$ and selecting $\Delta_j=\Delta_n^{\pm}\equiv \pm n(\tilde\nu-\Omega)-\tilde\omega$, we resonantly drive multiphoton Jaynes--Cummings interaction terms, while the rest of the terms in the expansion of the exponential term are off-resonant and rotating with a large frequency compared to its amplitude, i.e., $n|\Omega-\tilde\nu|\gg \epsilon_j/2$ (for zeroth order) where $n$ is the selected order of the interaction $\sigma^{\pm}a^{n}$. 
In this manner, performing these two approximations, one obtains:
\begin{align}
H_{\rm n}=\frac{\tilde\omega}{2}\sigma_z&+\tilde\nu\adaga+\sum_{j\in r}\frac{\epsilon_j(2\lambda)^{n_j}}{2\Omega^{n_j} n_j!}\left[ \sigma^+a^{n_j}+{\rm H.c.}\right]\\&+\sum_{j\in b}\frac{\epsilon_j(2\lambda)^{n_j}}{2\Omega^{n_j} n_j!}\left[ \sigma^+(-\adag)^{n_j}+{\rm H.c.}\right],
\end{align}
where $\Delta_{j\in r}= n_j(\tilde\nu-\Omega)-\tilde\omega$ and $\Delta_{j\in b}= -n_j(\tilde\nu-\Omega)-\tilde\omega$, which corresponds to Equation~(\ref{eq:Hn}). {The largest error committed in the previous approximation stems from the zeroth order in the expansion of the exponential. These contributions are of the form $\epsilon_j/2 (\sigma^+ e^{in_j(\Omega-\tilde\nu)t}+{\rm H.c.})$, which will produce a significant effect after a time $t\approx n_j(\Omega-\tilde\nu)/\epsilon^2_j$. For a single $n$-photon interaction term, population transfer occurs in a characteristic time $\tau_n=\sqrt{n!}(\Omega/2\lambda)^n/\epsilon_0$ (see Section~\ref{ss:nondiss}). Hence, we can provide a rough estimate for the duration of a correct simulation of the desired multiphoton Jaynes--Cummings model to be $t=k\tau_n$ with $k\approx (2\lambda/\Omega)^n n(\Omega-\tilde\nu)/(\epsilon_0\sqrt{n!})$.}



%
  
\end{document}